\documentclass[twocolumn,showpacs,preprintnumbers,amsmath,amssymb]{revtex4}

\usepackage{graphicx}
\usepackage{dcolumn}
\usepackage{bm}

\begin{document}

\preprint{ }

\title{All-optical generation and photoassociative probing of sodium Bose-Einstein condensates}

\author{R. Dumke, M. Johanning, E. Gomez, J. D. Weinstein, K. M. Jones and P. D. Lett}

\affiliation{Atomic Physics Division, National Institute of Standards and Technology, Gaithersburg, MD 20899-8424}

\date{\today}

\begin{abstract}
We demonsatrate an all optical technique to evaporatively produce sodium Bose-Einstein
condensates (BEC).  We use a crossed-dipole trap formed from light near 1 $\mu$m, and a simple ramp of the intensity to force evaporation.  In addition, we introduce photoassociation as diagnostic of the trap loading process, and show that it can be used to detect the onset of Bose-Einstein condensation.  Finally, we demonstrate the straightforward production of multiple traps with condensates using this technique, and that some control over the spinor state of the BEC is achieved by positioning the trap as well.
\end{abstract}

\pacs{03.75.Hh, 03.75.Be, 32.80.Pj, 39.20.+q}

\maketitle


\section{Introduction}

All-optical approaches to attaining Bose-Einstein condensation (BEC) allow the trapping of spin-zero species (which cannot be held in a magnetic trap), and multiple, arbitrary spin states at once.  In addition, the technique is convenient, avoiding large magnetic field coils that restrict optical access and allowing rapid evaporative cooling and condensation.  A number of variations on this technique have been used successfully to condense a variety of atomic species.  Here we present the first all-optical approach to evaporatively obtain BEC in sodium.  Additionally, we extend the technique to multiple traps in a configuration useful for atom optics experiments. Our stainless steel vacuum chamber incorporates an ion detector which we use to sensitively detect photoassociation transitions.  This signal is sensitive to the atomic density and provides a convenient way to optimize the loading of our dipole trap and BEC formation within the trap.

The first demonstration of evaporative cooling of a
sodium gas held in a crossed dipole trap \cite{Adams1995} was performed in 1995. The trap
was formed by a pair of beams from a Nd:YAG laser operating at 1.06
$\mu$m, detuned well red of the atomic resonance. In these studies the phase space density of
the gas was increased by a factor of 28, but did not reach the 
conditions required to Bose-condense the rather small sample used.  Later that year
BEC in sodium was first observed \cite{Davis1995} in a
trap that employed  magnetic confinement along with an optical
``plug'' to keep atoms away from a region of zero magnetic field.
While magnetic traps of many forms have since dominated 
BEC studies, interest in all-optical methods has continued as well.  

\begin{figure}

\includegraphics[width=\columnwidth]{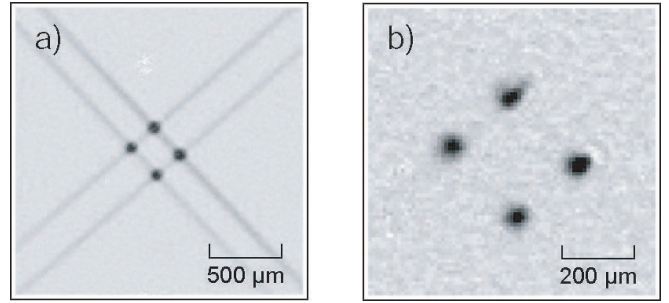}

\caption{\label{fourtraps} a) Absorption image of four crossed dipole
traps taken just after loading from a magneto-optical trap. This array was generated by recycling the trapping beams. The traps have a separation of $\approx$~250 $\mu$m.  Atoms loaded into the arms of the trap make them visible before evaproation.  b) Absorption image of the traps taken after evaporation, as described in the text, showing four BEC's.}
\end{figure}

In this paper we return to investigate evaporative cooling of sodium in a crossed-beam dipole
trap.  With sufficient initial loading of the dipole trap, we find that it is relatively straightforward to achieve
BEC with up to 2.5~$\times$~10$^5$ $^{23}$Na atoms by simply ramping down the trap laser intensity.
The dipole trapping beams give us flexibility in arranging the location of the condensate and allows us to create traps with different spin-component populations.  By recycling the beams, we construct multiple crossing regions with variable spacings.
For instance, we can easily arrange a square array of four optical traps, spaced by $\approx$~250~$\mu$m, and produce a condensate in each trap, as shown in Fig.~\ref{fourtraps}.  Such arrays of condensates can be useful in a number of atom optics and fundamental physics experiments.

\section{Background}

Producing BEC without the use of a magnetic trap was first demonstrated with $^{87}$Rb  in 2001 
\cite{Barrett2001}.  In this experiment the 
trap was formed by crossed, tightly-focussed beams from a CO$_2$ laser operating at a wavelength near 10~$\mu$m.  Since that time BEC in optical traps has been
observed with several other species as well, using traps and procedures of varying complexity.  $^{133}$Cs
\cite{Weber2003} was condensed in a combined CO$_2$/fiber laser dipole trap.  Cs is heavy and difficult to condense, requiring magnetic field gradients to cancel the force of gravity and bias fields to tune the scattering length by using a magnetic Feshbach resonance.  In this case weakly-focussed beams from the CO$_2$-laser trap the Cs atoms, and a tightly-focussed beam at 1064~nm is introduced to create a ``dimple" in the trap, where atoms can evaporate to higher density.
Cr was recently condensed \cite{Griesmaier2005} by an evaporative cooling sequence which begins in a magnetic trap, and is completed after transferring the atoms to a crossed-dipole optical trap.  Such traps are not restricted to the infrared:  $^{174}$Yb, with its strongest transition near 400 nm, was condensed in a crossed-dipole trap \cite{Takasu2003b} formed from a 532 nm laser.   The technique is also applicable to fermions, and evaporative cooling in a single beam CO$_2$-laser optical trap
has also been used to produce a degenerate Fermi gas of  $^6$Li
atoms \cite{Granade2002}.  

Many variations of the loading sequence, including the ramping of laser intensity and magnetic field, as well as dynamic changes in the optical system, have proven successful.  Many of these complexities are dictated by the specifics of the atomic species that are used, especially their collision properties.
In comparison with many of these previous experiments, sodium BEC can be achieved in a relatively simple sequence.
No magnetic field gradient is required to compensate for gravity, and we use 1.07~$\mu$m light from a fiber laser, which does not require the special window materials that are needed in the far infrared.  Furthermore, we use a simple intensity ramp with no moving optics, resulting in a relatively uncomplicated and robust procedure.

We begin our experiment with atoms accumulated 
in a magneto-optical trap (MOT) fed with a zero-crossing Zeeman slower.  Our MOT is of order 2~mm in diameter, while the optical trap has a 1/$e^2$ waist of
61~$\pm$~3~$\mu$m, resulting in a poor geometric overlap for loading.  As
discussed in Ref.~\cite{Kuppens2000} the loading process is somewhat
complex and various schemes have been developed for improving it.
To attain BEC in our experiments we have maintained a fixed trap size
and concentrated on optimizing the loading by adjusting the way in which we
ramp down the MOT light and magnetic field.

\begin{figure}

\includegraphics[width=\columnwidth]{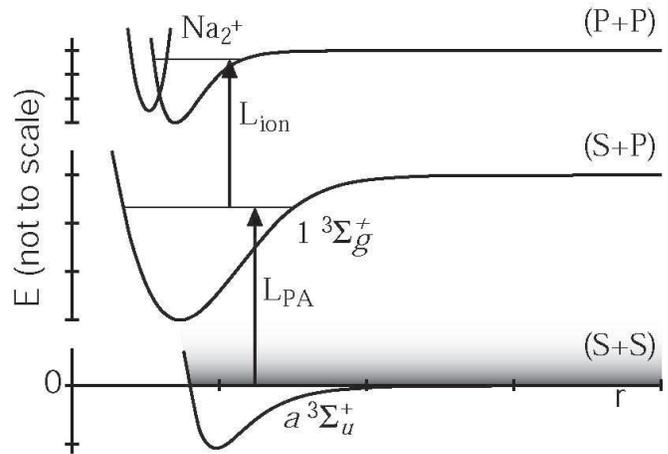}

\caption{\label{PAsketch}  Sketch of the relevant Na$_2$ molecular
potentials and coupling lasers. The photoassociation laser L$_{PA}$
couples colliding atoms with energy E to form an excited-state
molecule. The excited state is resonantly excited to an autoionizing 
state by L$_{ion}$, resulting in an ion signal proportional to the
excited state population.}
\end{figure}

To monitor the buildup of the atomic density in the optical trap we introduce a
photoassociation technique. 
Photoassociation (PA) is the process
where two colliding atoms absorb a photon and form an electronically excited
diatomic molecule \cite{Weiner1999,Jones2006} (see Fig. \ref{PAsketch}).  Photoassociation has been used to produce high-resolution spectra of the molecular states that are accessible in this way.  
The photoassociation transition can be detected in a number of ways, the two most common are trap loss and ionization.  When a molecule is formed two atoms are removed from the trap, and a simple detection of atom number (by absorption or fluorescence), as a function of frequency, can be used to locate a PA transition.  If the excited molecule can be further excited to an autoionizing state, or to an ionization continuum, the presence of an ion signal will indicate the PA transition.
On the other hand, the technique can also be used as a probe for a number of effects, such as wavefunction distortion by magnetic \cite{Courteille1998} or optical Feshbach \cite{Fatemi2000} 
resonances, or, as in this case, as a sensitive probe of the atomic density.  Many of the low-lying states of Na$_2$ have been studied using photoassociation techniques, and in the current experiments we take advantage of a photoassociation-ionization transition that we have used in previous spectroscopy experiments \cite{deAraujo2003}.  

We demonstrate that we can monitor the evolution of the trap density in our experiment using a PA ionization signal produced by two auxiliary lasers.  One laser is tuned to a
known molecular transition and the second is tuned to promote the excited
molecule to a high-lying autoionizing state.  The result is the
production of Na$_2^+$ ions, which we detect with a microchannel
plate.  For the present purposes the important feature of the PA
signal is that it depends on the square of the atomic density, and thus the signal is most sensitive to atoms in the dipole trap rather than those in the diffuse cloud of atoms around it, even though there are many more atoms in this cloud.   Within the trap it is most sensitive to atoms in the lowest vibrational states of the trap, where the density is highest.

\section{Summary of Trap Loading and Evaporation Procedure}

\begin{figure*}

\includegraphics{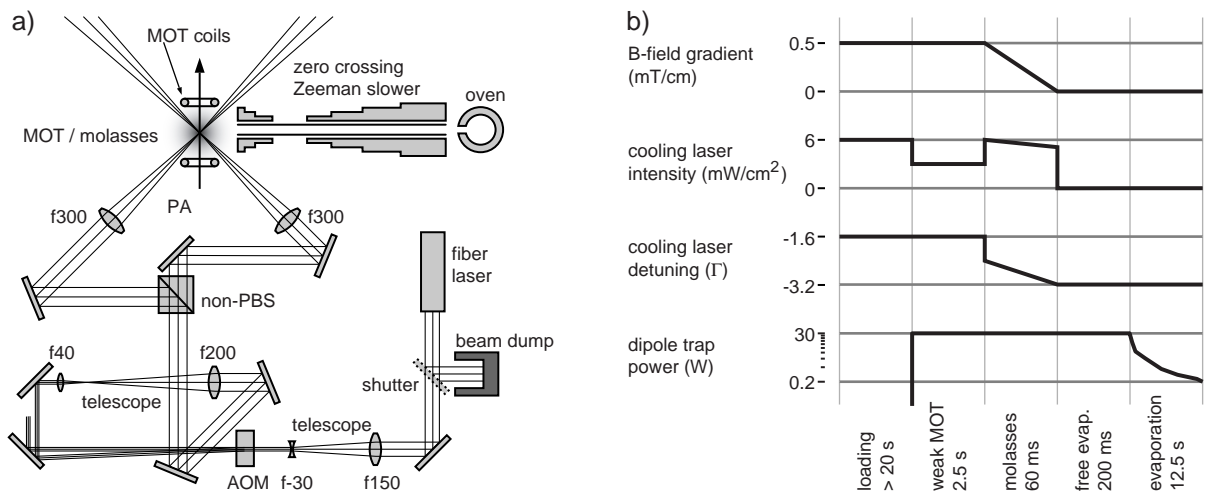}
\caption{\label{experiment} a) Sketch of the experimental
setup.  
b)  Timing sequence of the experiment indicating the MOT magnetic field gradient, MOT/cooling laser detuning and intensity, and the total power in the dipole trap beams. There are four stages to 
the loading of the trap.  A strong MOT period is used for rapid loading.  The weak MOT period then reduces the temperature of the MOT.  These are followed by an optimized ramped-molasses phase, and a free evaporation phase, where the intensity of the dipole laser is
held constant. Finally, there is a forced evaporation phase where the intensity of
the dipole laser is successively reduced.}
\end{figure*}

The experimental setup and timing sequence are shown in Fig.~\ref{experiment}a and  Fig.~\ref{experiment}b, respectively.
Our approach to attaining BEC proceeds as follows:  
The dipole trap is loaded from a standard Zeeman-slower-fed, six-beam,
dark-spot, magneto-optical trap (MOT). 
This produces, after a loading time of about 8~s, a sample of 7$\times 10^9$ $^{23}$Na atoms at about 300~$\mu$K, in a volume of $\approx$(2 mm)$^3$. 
A single, traveling-wave, annular repumping beam, tuned to the 3S F=1 $\rightarrow$ 3P$_{3/2}$ F=2 transition, is used along with the cooling light, tuned to the red of the 3S F=2 $\rightarrow$ 3P$_{3/2}$ F=3 cycling transition, to form the dark-spot MOT.  
Nearly all of the atoms are thus in the F=1 hyperfine state. This
reduces hyperfine-changing collisions in the dipole trap and is required
for the application of further cooling techniques.  
The dipole trap is formed by crossing two focussed beams in the MOT volume, as described below, and is turned on during the final 2.5 seconds 
of the MOT phase.  At this time the intensity of the cooling light
is reduced from 6 mW/cm$^2$ to 3 mW/cm$^2$ in each beam, producing a weak MOT
to further cool the atomic sample.
Loading the dipole trap directly from the MOT at this point,
gives a trapped atom number of only 3$\times 10^4$ (measured by
absorption imaging 100 ms after the MOT phase). 

To increase the
number of atoms loaded into the dipole trap a 60 ms transitional ``molasses" phase is applied as the MOT is gradually turned off.
During this time the magnetic field gradient from the MOT coils is lowered from 0.5~mT/cm to 0~mT/cm and the detuning of the cooling light is increased from 24 MHz
to 32~MHz (measured red from the 3S$_{1/2}$ F=2 $\rightarrow$ 3P$_{3/2}$ F=3
cooling transition at 16973 cm$^{-1}$). The repumping beam during this period continues to be a single, annular, traveling-wave beam.  The ramp starts from a detuned strong MOT condition that at first compresses the atomic cloud.  As the ramp continues the temperature of the cloud is decreased.
The final temperature of the cloud after this
phase is 50~$\mu$K. The currents in a set of magnetic-field-nulling coils are adjusted in order to achieve the lowest temperature at this phase.  After this 60 ms molasses phase all near-resonant laser beams
are extinguished by acousto-optical modulators (AOMs), followed by mechanical shutters.
This loading process determines the initial phase space density, and thus,
whether or not it is finally possible to reach quantum degeneracy. 

The molasses phase is followed by a 300 ms period of free evaporation, 
during which the dipole trapping beams are held at constant power.  
Finally, we have a period of forced evaporation during which both dipole trapping beams are ramped down in power, simultaneously, over a time of 12.5 s.  The timing sequence of the different experimental steps is shown in Fig.~\ref{experiment}b.  The details of the experiment, and the diagnostics applied to optimize this sequence are discussed in the next section.

\section{Experimental Procedure and Photoassociation Diagnostics}

The trapping beams are produced by a commercial Yb-doped fiber laser which
can generate up to 100~W of 1.07~$\mu$m light. The line width of this laser is
$\approx$~3~nm and the polarization is random. (The
polarization changes quickly on the time scale of the typical oscillation
periods of the atoms in the dipole trap.) In practice we operate the
laser at $\lesssim$~40~W, since it shows excess intensity noise at higher
powers. 
The trapping light passes through one AOM. The
first diffracted order from the AOM goes to the experiment and the intensity of this beam is
controlled by the rf-drive power. The beams are divided by a
non-polarizing beam-splitter cube and focussed into the vacuum
chamber with 300 mm focal length lenses. Note that we do not need to polarize the
beams or introduce frequency shifts to wash out interference effects. The
angle between the beams inside the chamber is $\approx90^{\circ}$ and the beam
waist in the crossing region is $\approx$~61~$\mu$m (1/$e^2$ radius). 
The beam size is obtained from the trap oscillation frequency, laser power, and calculated Stark shift.
The plane of the crossed beams is perpendicular to gravity.  Each beam has a
maximum power of 14~W  at the trap location, corresponding to an initial potential depth of $U/ k_B = 510~\mu$K (where k$_B$ is the Boltzman constant). 

To analyze the properties of the atoms in the dipole trap, we can use
absorption imaging. The atoms are released from the trap by turning off the trapping laser in order to minimize effects due to ac-Stark shifts caused by the dipole trap. 
The number and temperature of the atoms are extracted using
time-of-flight (TOF) techniques.
For short times after the end of the molasses phase, the absorption signal is dominated by the expanding MOT cloud and it is necessary to
allow the non-trapped fraction of the atoms to fall away before we can cleanly observe the atoms in the dipole trap.  500~ms after switching off all resonant light we
trap $\approx10^7$ atoms in the whole dipole potential, and 2~$\times~10^6$ atoms in the crossed region, with a temperature
of 50~$\mu$K. This results in an initial peak atom density of 9~$\times~10^{13}$~cm$^{-3}$.  The density is derived from the measured
temperature, atom number, and trap oscillation frequency, and by assuming that the Gaussian potential can be described by a harmonic approximation at the bottom of the trap.

To measure the trap oscillation frequency we parametrically excite the breathing mode of the trap.  
First we evaporatively cool the atoms, proceeding 3 s into the usual intensity ramp of the dipole trap (discussed below).   We then modulate the power of the dipole trap potential sinusoidally and measure trap loss as a function of modulation frequency using time-of-flight imaging. 
 The mean oscillation frequency of the trap during the free evaporation stage is 
1.7~$\pm$~0.2~kHz. 
Trap oscillation frequencies for other laser powers are derived from this by scaling to the square root of the measured power in the trap beams.
We obtain the trap depth and the waist size from the oscillation frequency and the power in the trapping beams, again making a harmonic approximation \cite{Friebel1998}. 

To probe the evolution of the atom cloud during the molasses and evaporation stages, we employ a new technique which allows us to directly monitor
the density of the atoms inside the trap.  
The density
is probed by a PA laser which resonantly produces bound excited state molecules  (in the $J=0$ rotational level of the $0_g^-$ component of a vibrational level in the $(1)^3\Sigma_g^+$ (S+P$_{3/2}$) potential \cite{Xu2005}).  The
laser frequency for this transition is 16845.16 cm$^{-1}$.  The molecules which are
formed are then excited by an additional laser (at 17099.95 cm$^{-1}$) to an autoionizing level 
(see Fig. \ref{PAsketch}). The ions that are
produced are then efficiently detected by a microchannel plate.  This ion signal is proportional
to the volume integral over $n^2$, where $n$ indicates the atom
density.  At the temperatures under consideration here, this $s$-wave PA transition is in the Wigner Law regime and the PA signal is not directly dependent on the temperature (average collision energy) of the atoms, but simply depends on the density (as discussed in \cite{Jones2006}).

\begin{figure}

\includegraphics[width=\columnwidth]{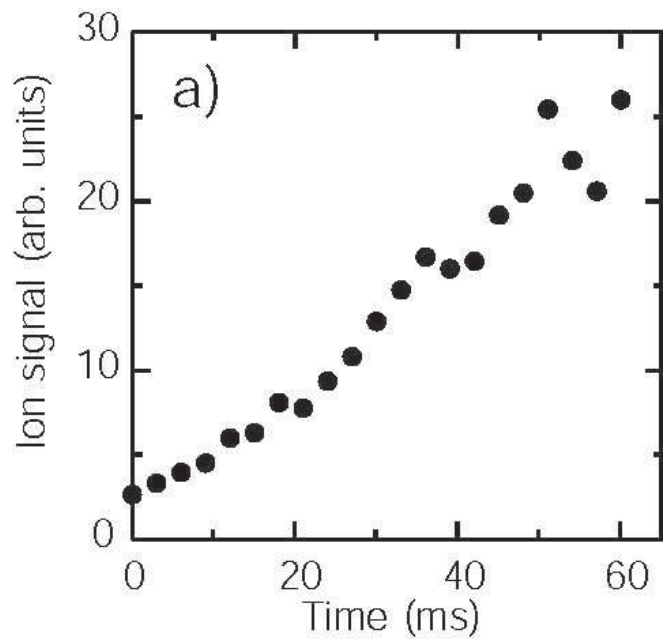}

\caption{\label{MOTload} 
The Na$_2^+$ ion signal due to
photoassociation (PA), as explained in the text, which is used to
optimize the trap loading sequence.  The MOT laser tuning is shifted and the
magnetic field ramped down over 60 ms. This scan shows the resulting ion signal
when the ramping has been optimized. 
}
\end{figure}

In sodium the molecular energy levels are such that in a MOT a steady ion signal is produced by the trapping beams themselves, so it is necessary to turn them off to see the PA signal.
The PA and ionizing lasers are then turned on (4~mW and 5~mW, respectively, in an approximately 100~$\mu$m spot size).
The ion signal that is produced can be used to optimize the cooling process and the loading into the trap.
Figure \ref{MOTload} shows the ion signal, measured in this way (integrated over a 10~ms period), during our optimized
60~ms molasses ramp. The density increases continuously until the end of the 60~ms, at which point the magnetic field is zero.  During the last few milliseconds of this stage the magnetic field at the location of the dipole trap is so small that the conditions are characteristic of a true optical molasses, and the final temperature is not improved by extending the molasses period beyond the end of the magnetic field ramp.  

After the molasses phase we turn off all of the cooling light beams
and wait before starting the forced evaporation ramp. The evolution of
the density during this first 300 ms free evaporation period helps  to reach the
extremely high phase space density needed for a successful forced
evaporation.  

\begin{figure}

\includegraphics[width=\columnwidth]{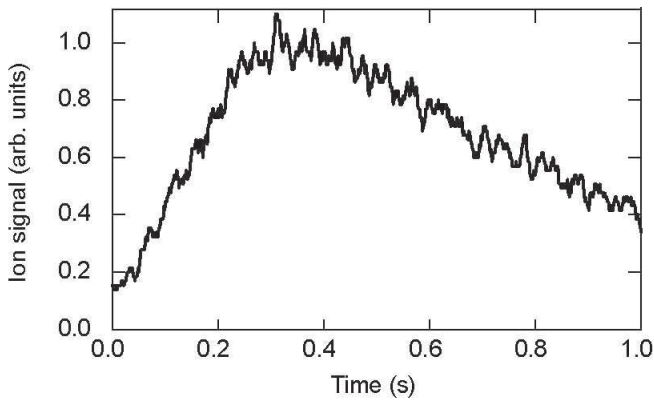}

\caption{\label{freeevap} Ion signal during the first one second of evaporation with constant trap intensity.  During this stage all additional cooling light is turned off and the PA and ionizing laser powers are lower than in Fig. \ref{MOTload}. The ion
signal (integrated for 10~ms periods) indicates the density evolution during this free evaporation
stage. The density increases during this phase while atoms are lost, which
indicates that during this time the atomic cloud is rapidly cooling.}
\end{figure}

During the free evaporation phase (where the dipole laser power is held
constant), we can again probe the time evolution of the density of the atoms in the
trap by using our PA technique.
We are continuously producing and ionizing molecules during this time, however, 
a 1~s duration PA pulse (at a power of 340~$\mu$W) reduces the atom number by only 5$\%$ and represents 
a small perturbation on the trapped atom density dynamics.  In Fig. \ref{freeevap} the
ion signal shows that the density begins to increase during this trap
holding phase.  
The evaporation
process causes atom loss and is not compensated by any loading during this time.
The remaining atoms rethermalize at higher density.
The increase in density slows down with time and, finally, if held too long, the
loss out of the trap becomes the dominant process and the density starts
to decrease.  This can be seen for times later than about 300~ms in Fig.~\ref{freeevap}.  The trap lifetime under these conditions, as seen in the figure, is fairly short and is possibly limited by beam-pointing instabilities induced by thermal effects (due to
the high RF power in the AOM which controls the power into the
trapping beams).  Adding feedback to the AOM to stabilize the laser intensity does not improve the lifetime.  Under other conditions longer lifetimes are obtained, as discussed below.

The initial phase space density in the trap,
just 500 ms 
after the molasses phase (or 200~ms after the free evaporation phase), is 1~$\times~10^{-2}$. Using the corresponding
value for the density and the three-body rate constant from \cite{Stamper1998} (K$_3 = 1.1~\times~10^{-30}$ cm$^6$s$^{-1}$, measured for the F=1, m$_F$=-1 state, with a density of 10$^{14}$~cm$^{-3}$), we can infer the three-body loss rate in our trap to be $\approx~$ 0.01~s$^{-1}$. 
It is thus possible that the three-body losses contribute to limiting the density achievable
by free evaporation. The estimated elastic
collision rate is initially $\approx~$2~kHz. During the optimized loading
ramp and after free evaporation the density increases by about a
factor of 300, from 3~$\times$~10$^{11}$ in the MOT, to 9~$\times$~10$^{13}$~cm$^{-3}$ in the dipole trap.

\begin{figure}

\includegraphics[width=\columnwidth]{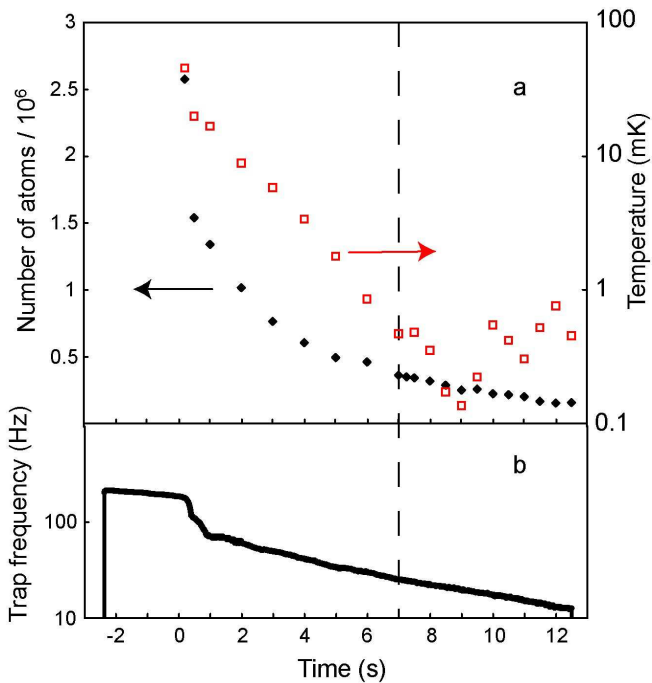}
\caption{\label{NTramp} 
a) The evolution of the temperature of the thermal fraction (squares) 
and the trapped atom number (diamonds) during
the evaporation ramp. The temperature of the thermal fraction changes linearly with the
potential depth until approximately the point where a condensate starts to form. The zero of time is the beginning of the forced evaporation ramp, after 300 ms of free evaporation with a static trap depth.
b)~The evaporation ramp (on a log scale) of the trap frequency versus time.   The dashed vertical line indicates the time of the onset of BEC. }
\end{figure}

At the time when free evaporation gives the maximum density we start a period of forced
evaporation by simply lowering the power simultaneously in
both dipole trap beams.  The most energetic atoms escape and the
remaining atoms rethermalize at a lower temperature with a higher
phase space density.  The rethermalization rate given by the elastic
collisions slows down as the trap potential (and therefore the 
oscillation frequency) is lowered.  At the end of our evaporation
ramp we calculate a factor of 25 lower collision rate than at the beginning.  Our ramp is divided into 7 segments during each of which the intensity is dropped, approximately linearly, by a factor of two (see Fig.  \ref{NTramp}b).  (The ramp function fed to the electronics is linear, however, nonlinearities in the rf drive electronics and the AOM response lead to some curvature in the actual ramp segments, which are not corrected for.)  The duration of each segment is optimized
experimentally to give the largest phase space density.  The temperature at the end of each segment is determined by the trap depth at the end of that segment.  Our ramp is 12.5~s long
and approximately follows the theoretically optimized curve $U(t)\propto
1/(1+t/\tau)^\beta$ derived in Ref. \cite{OHara2001b}.  Our ramp function, except for the earliest times, is actually closer to an exponential, with a decay constant of 2.6 s.  Making a fit to the above function, however, results in 
values of $\tau$ = 22 s and $\beta$ = 9, giving an acceptable fit, again except for the earliest part of the ramp.  This value of $\beta$ is above the maximum that would be allowed under the assumptions of  Ref. \cite{OHara2001b}, where $\eta = U/k_BT$, is kept constant.

\begin{figure}

\includegraphics[width=\columnwidth]{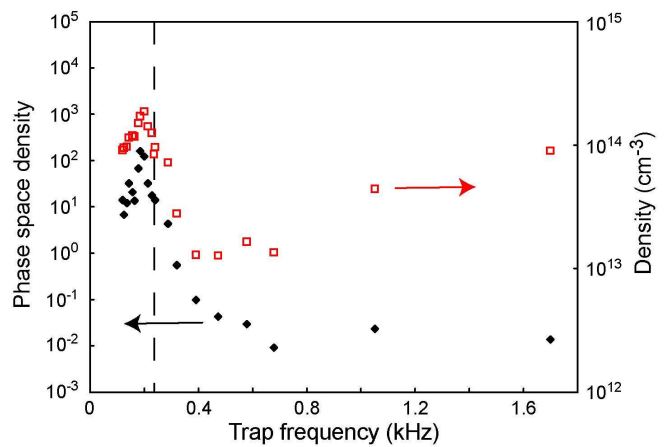}

\caption{\label{nphase} 
Density evolution in the trap during the forced evaporation ramp (squares, right vertical axis) as a function of trap oscillation frequency.  The peak phase space density as a function of the trap oscillation frequency (diamonds, left vertical axis).  The dashed vertical line indicates the onset of BEC and to the right of this line the condensate fraction is less than 1 $\%$.}
\end{figure}

Figure \ref{NTramp} shows the decrease of the atom
number and temperature with the lowering of the trap oscillation frequency.  The temperature
decreases monotonically, roughly as a constant fraction $\eta$~$\approx$~10 of the potential depth \cite{OHara2001b}. Figure \ref{nphase} shows the evolution of the peak density, derived from the data in Fig.~\ref{NTramp} and assuming a harmonic potential. (Once BEC begins we fit the atom distribution to a sum of two components: a broad thermal distribution and a narrow distribution representing the condensate fraction.  The temperature is extracted from the fit to the thermal fraction and the peak density from a sum of the densities of the thermal and the condensed components.)    The
density decreases at first, due to the loss of atoms and the decreasing confinement as the trap is weakened.  The density then rises as the condensate starts to form. 
As the intensity, and thus the trap frequency, is lowered, the temperature also decreases, as shown in Fig.~\ref{NTramp}. Taking all the parameters into account one sees that
the resulting phase space density increases as the
evaporation proceeds, as seen in Fig.~\ref{nphase}. The speed of this increase is sensitive to the initial
conditions.  
At the lowest trap frequencies most of the atoms are already condensed and the density and phase space densities begin to go down as the trap potential is further expanded. 
At the end of the 12.5 s intensity ramp we observe a condensate with a
typical number of 1.5~$\times$~10$^5$ atoms, or about 80$\%$ of the atoms. The final trap depth is 2.9~$\mu$K and the oscillation frequency is 120~Hz with 80~mW of power in each trapping beam. The peak density is 9~$\times$~10$^{13}$~cm$^{-3}$, and the observed condensate lifetime is 18~s.  Most likely this lifetime is limited by noise on the trapping potential.  The background gas
pressure is 4~$\times$~10$^{-9}$~Pa (3~$\times$~10$^{-11}$~torr), measured by an ionization gauge.
This would correspond to a vacuum-limited lifetime of some hundreds of seconds, however the 3-body decay rate given above would limit the lifetime to closer to 100 s.

\begin{figure}

\includegraphics[width=\columnwidth]{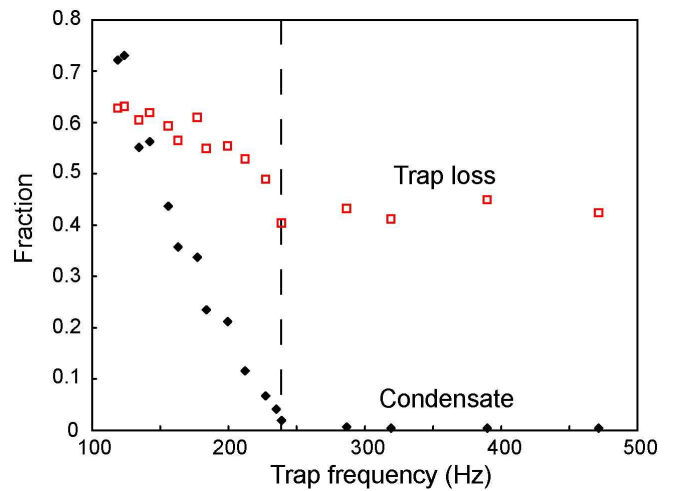}

\caption{\label{fracloss} 
Condensate fraction of the trapped atoms (diamonds)
and fractional trap loss due to a 10~ms, on-resonance photoassociation pulse, as explained in the text (squares), as functions of the trap oscillation frequency. 
The dashed vertical line indicates the onset of BEC.}
\end{figure}

A straightforward investigation of the onset of BEC can be
made by probing the atoms on a photoassociation transition and
measuring the PA rate by either the ion signal, as
discussed above, or the loss of the atoms out of the trap, as shown
in Fig.~\ref{fracloss}.  Here the PA laser intensity is much higher than in Fig.~\ref{freeevap}. 
We can observe the onset of BEC using the PA technique as it detects the steep rise in density
that accompanies the phase transition.
PA trap loss measurements sometimes result in higher signal-to-noise than the ion signal discussed above, and can also be used for such measurements, albeit destructively. 

Measurements of the fractional trap loss are shown in Fig.~\ref{fracloss} and are produced by illuminating the trap with a
laser resonant with the PA transition discussed above.
The cloud is illuminated for a time of 10~ms and contains 4~ mW in a $\approx$~100~$\mu$m spot.  
The remaining atom
number is compared with the atom number when the PA beam is tuned
off resonance to give the fractional trap loss measurements plotted in the figure.  Near the end of the evaporation ramp the trap depth is only
$\approx$~2.9~$\mu$K and the probe beam, which is detuned red from the atomic
resonance, significantly distorts the trap potential by inducing an
additional dipole force on the atoms. To prevent this effect a
laser, blue detuned from the resonance ($\approx$~5~mW), is added to counteract this dipole force.
Fig.~\ref{fracloss} shows that when the
condensate emerges the fractional trap loss increases rapidly
and can serve as an indicator of the presence of the condensate. 

\begin{figure}

\includegraphics[width=\columnwidth]{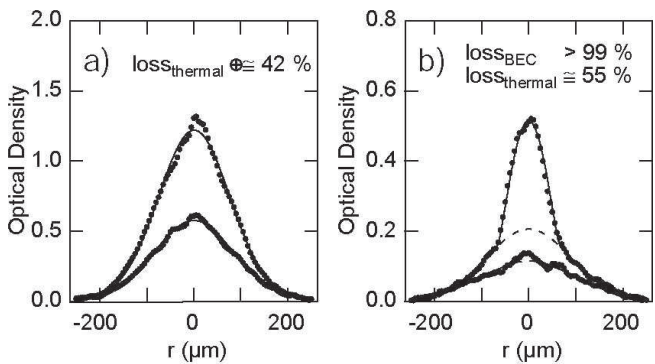}

\caption{\label{PAloss} Line profiles taken from time-of-flight pictures of an
atomic ensemble.  The upper traces are taken with the photoassociation (PA) beam tuned off resonance
and the lower profiles after applying the
PA beam on resonance.  These profiles are produced by finding the center of a time-of-flight picture of an atom cloud and azimuthally averaging over $180^{\circ}$.  The two independent half-circles of data, one used for negative r, allow for better signal-to-noise, while the asymmetry in r permits some judgment of the noise level. 
a) Time-of-flight profiles with a thermal kinetic energy distribution and fit.
b) Time-of-flight profiles of an atomic ensemble
with an initial condensed fraction of 41$\%$. The lines through the data are two-component (thermal/BEC) fits to the data.  The dashed lines show the thermal contribution.
}
\end{figure}

To understand Fig.~\ref{fracloss} in more detail it is
essential to know which parts of the atomic distribution are
contributing to the PA signal.  Figure \ref{PAloss}
shows line profiles of time-of-flight pictures of the atom
distribution, taken with and without the PA beam applied (with the same intensity and duration as above).  
Figure \ref{PAloss}a was taken when there was only
a thermal cloud present, and Fig.~\ref{PAloss}b was taken after the appearance of a condensed
fraction within the thermal cloud.  The upper profile in both of the
plots show the profile when the PA beam was applied but tuned off-resonance,
while the lower profiles show which atoms remain after the PA beam
tuned to resonance was applied. In both cases one sees that after the PA pulse
a large fraction of the coldest atoms are missing.  This is because the
coldest atoms are mostly held near the trap center, where the density is
highest.  When a fraction of the atoms are condensed,
as in Fig.~\ref{PAloss}b, the condensed part 
contributes the most to the PA signal because this part
of the atomic distribution has the highest density - even though the
PA rate coefficient for a condensate is lower by a factor of two than that of
a thermal ensemble \cite{Harber2002,Jones2006}.

\section{Discussion}

As shown in Fig. \ref{fourtraps} we can quite easily create a number of condensates by recycling the trap beams, as long as the crossing regions stay within the volume of the MOT.  The trapping beams are brought into the chamber through 300 mm lenses and are diverging as they exit the vacuum chamber.  Each exiting beam is then recollimated by another 300 mm lens, and retroreflected back through the same lens on a small angle, to intersect the orthogonally-propagating beams.  This simple construction can produce an array of four traps.  Such an array of condensates immediately suggests that one might use them in a variety of atom-optics experiments.  One can, for instance, imagine launching one condensate into another by scanning the retro-reflecting mirror.  With more sophisticated optics one can think of creating a regular lattice array of condensates in three dimensions, with spacings on a scale of about 100 $\mu$m.  

\begin{figure}

\includegraphics[width=\columnwidth]{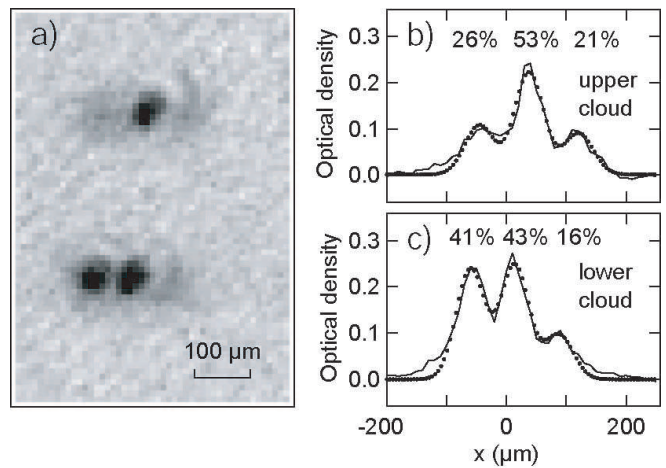}

\caption{\label{spins} a) Absorption images after a time-of-flight 
of two condensates generated by retro-reflecting one trapping
beam as described in the text. During the time-of-flight a magnetic field gradient is
applied, separating the m$_F$ levels.   Line trace of the distribution of atoms between the m$_F = 0,\pm1$ states for the upper (b), and lower (c) cloud.  The distribution depends on the position of the trap in the MOT.  The dotted lines represent a fit to the sum of three displaced Gaussians, from which the relative populations are extracted.}
\end{figure}

Optical traps can produce spinor condensates, as demonstrated in Ref.~\cite{Barrett2001}.  By applying a magnetic field gradient during time-of-flight imaging one can separate and observe the different magnetic spin components.  We
can also create a mixture between the three available spin states, m$_F = 0,~\pm1$.  In Fig.
\ref{spins} we have recycled and refocused only one of the trapping
beams to generated two traps. By applying the same 
evaporation ramp to all of the beams we generate two condensates at the same
time. We then apply a magnetic field gradient with our MOT coils and, in this case, we find that the population
of the magnetic spin components is different for the two traps.  In fact, the spin distributions map to the  different trap positions and allows for almost pure spin states under some conditions. By successively changing the crossing position and having a
little bit luck, one is able to obtain most of the population in a single
magnetic sublevel without having to apply any additional optical pumping.
The polarization of the sample is apparently due to the local magnetic field and laser imbalances in the MOT and molasses phases, which produce varying m$_F$ distributions in different spatial locations, which are then preserved in the trap.

The technique that we describe here represents our first effort and could be improved upon by, amongst other things, stabilizing the laser intensity at higher powers with an AOM.  Problems due to pointing instabilities could also be improved by better heat-sinking of the AOM to reduce temperature fluctuations caused by changes in rf powers to the AOM.  The beam spot size has not been optimized in these experiments, and it is still to be determined if we can trap more atoms and produce larger condensates by using a larger spot size (taking advantage of the factor of 2-3 more power that we have available). Eventually, at large trap sizes the evaporation to BEC will be stalled by the lack of density or limited by background gas scattering as we require longer evaporation times.

In this paper we have demonstrated a straightforward, all-optical method of production of sodium Bose-Einstein condensates. With some variation in parameters, this approach could be used for other species as well.    In addition, we have introduced a new
technique for probing the density evolution in the dipole trap
during the evaporation process using photoassociation.  This tool enables us to
optimize the production of a Bose-Einstein condensate in a simple manner.  Under some conditions it also allows for non-destructive monitoring of the condensate evolution.
The fixed-focus crossed-dipole trap technique is capable of generating multiple
condensates simultaneously, which can be connected with each other via the dipole trap beams.  The simple optical system allows for a great deal of flexibility in constructing experiments, and also provides good optical access to the system.  The ability to produce multiple independent condensates
suggests a number of interesting atom optics experiments
where, for instance, one could collide or channel condensates in the waveguides provided by the trapping beams.


\end{document}